\documentclass{article}
\usepackage[T1]{fontenc}
\usepackage{spconf,amsmath,amssymb,graphicx}
\usepackage{booktabs}
\usepackage{multirow}
\usepackage[table]{xcolor}
\usepackage{adjustbox}
\usepackage{url}
\usepackage{balance}
\usepackage{hyperref}
\hypersetup{colorlinks=true,citecolor=blue,linkcolor=blue,urlcolor=blue}
\definecolor{laglightblue}{HTML}{E8F3F8}
\newcommand{\method}{RefTAG}
\newcommand{\recot}{Re-CoT}
\title{RefTAG: Reflection-aware Temporal Audio Grounding in Large Audio Language Models}
\name{\shortstack{Yuhang Dai$^{1}$, Xin Shu$^{2}$, Zhennan Lin$^{1}$, Mingchen Shao$^{1}$,\\ Chengyou Wang$^{1}$, Xiangang Li$^{2}$, Jianwei Yu$^{2}$, Lei Xie$^{1,*}$}%
\thanks{$^{*}$ Corresponding author.}}
\address{
    $^1$ Audio, Speech and Language Processing Group (ASLP@NPU), School of Computer Science \\
    Northwestern Polytechnical University \\
    $^2$ Independent Researcher
}

\begin{document}
\ninept
\maketitle

\begin{abstract}
Temporal audio grounding requires locating every occurrence of a queried target, making both occurrence counts and interval boundaries essential. We propose \method{}, a post-training framework for verifiable temporal self-correction. Its Reflection Chain-of-Thought (\recot) protocol combines draft intervals, discrete count and boundary actions, and revised intervals in one response. Supervised fine-tuning on these trajectories (\recot-SFT) teaches the protocol; reinforcement learning (RL) jointly rewards final localization and revision outcomes. Across 8,228 queries, \recot-SFT improves query-weighted mIoU over final-answer SFT by 1.74 percentage points. RL further improves TAG-Bench multi-target mIoU by 0.40 points and R@0.5 by 0.86 points. The final model achieves 38.50\% TAG-Bench mIoU and 51.01\% mean mIoU across three datasets, outperforming the evaluated open-source MOSS-Audio and FireRedAudio models and the commercial Gemini-3.1-Pro on both measures. These results demonstrate that explicitly coupling count- and boundary-aware revision with joint reward optimization provides an effective and verifiable path to temporal audio grounding.
\end{abstract}


\section{Introduction}
\label{sec:intro}
Temporal audio grounding (TAG) connects natural-language descriptions to the intervals in which the described content occurs~\cite{xu2021tag}. Unlike clip-level recognition, it requires semantic identification and temporal evidence: when a target starts, when it ends, and whether it recurs. Under TAG-Bench's all-occurrence setting~\cite{dai2026tagbench}, one plausible interval is insufficient for a target appearing in several disjoint regions. Grounding therefore requires a variable-size interval set with accurate coverage and boundaries, providing evidence that can be inspected against the recording.

Large audio-language models (LALMs), including Audio Flamingo 3~\cite{goel2025af3} and Qwen3-Omni~\cite{xu2025qwen3omni}, support natural-language audio interaction. Temporal extensions address specific bottlenecks: TimeAudio introduces temporal representations~\cite{wang2026timeaudio}, TimePro-RL combines time prompts with temporal-reward optimization~\cite{shi2026timepro}, and LAT-Audio retrieves local evidence through iterative tool use~\cite{shao2026lataudio}. These methods improve temporal information access or localization supervision. Our focus is the prediction process that follows: identifying and correcting errors in an already generated interval set.

Temporal correction has two coupled difficulties. First, occurrence and boundary errors require different interventions: shifting a boundary does not recover an omitted occurrence, while adding intervals does not repair inaccurate endpoints. TAG-Bench reports under-counting on recurring targets~\cite{dai2026tagbench}; event-presence analyses also find reduced reliability in scenes with more events~\cite{lee2026multievent}. Second, a fluent critique does not measure localization improvement. Final-answer supervision alone cannot distinguish useful revision from preservation of an accurate draft. Two incorrect predictions can also differ in overlap, count, and boundary accuracy. Reflection must therefore specify the correction and quantify its effect.

These observations motivate reflection over interval predictions rather than unrestricted explanations. Building on reflection-aware training~\cite{wan2025srpo}, we expose a draft, diagnose count and boundary errors, and evaluate the revision against the same temporal criterion. This connects the intermediate response to the grounding objective without a free-text reflection judge.

We propose \method{} with the Reflection Chain-of-Thought (\recot) protocol. Each response contains draft intervals, discrete count and boundary actions, and revised intervals. Supervised fine-tuning (SFT) teaches this structure; reflection-aware Group Relative Policy Optimization (GRPO) rewards final quality and revision outcomes, including preservation of accurate drafts. A shared quality function combines coverage, interval matching, midpoint accuracy, and count agreement. The backbone is unchanged, and inference requires no annotations.

Our contributions are the structured correction protocol and the interval-based reflection objective. Experiments separate final accuracy, draft informativeness, and revision gain, then test reward mixing and individual quality components. Together, these comparisons establish the protocol's grounding gains and identify the additional benefits of reflection-aware reinforcement learning.

\begin{figure*}[t]
  \centering
  \includegraphics[clip,trim=2.8cm 21.5cm 2.6cm 3.8cm,width=\textwidth]{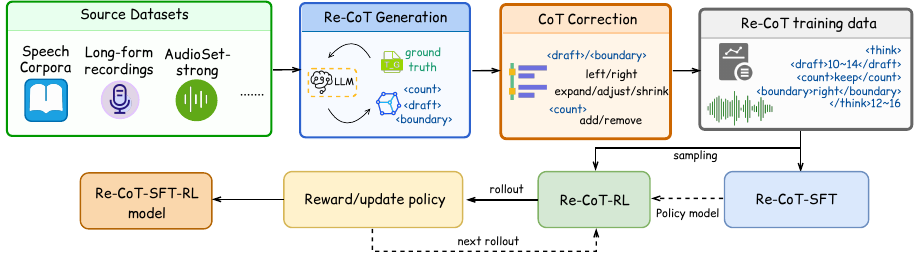}
  \vspace{-0.5cm}
  \caption{\method{} data construction and training. Ground-truth-conditioned LLM generation and rule-based correction produce balanced \recot-SFT trajectories. \recot-RL generates fresh rollouts on 5K SFT examples. Ground truth guides data construction and rewards, never inference.}
  \label{fig:pipeline}
  \vspace{-0.5cm}
\end{figure*}

\section{Related Work}
\label{sec:related}
\noindent\textbf{Audio-text alignment and supervision.}
Early TAG predicts event boundaries from similarities between frame-level acoustic embeddings and a phrase embedding~\cite{xu2021tag}. Weakly supervised TAG learns these correspondences from clip-caption pairs using phrase-level matching, negative sampling, and pseudo-label supervision~\cite{xu2024wstag}. AudioTime constructs synthetic audio-caption pairs with controlled timestamps, durations, ordering, and occurrence frequencies~\cite{xie2025audiotime}; TACOS annotates temporal regions in recordings and uses frame-wise contrastive learning~\cite{primus2025tacos}. These approaches improve the alignment signal through annotation or representation learning. \method{} addresses a different training target: diagnosing and revising the interval set generated by an LALM.

\noindent\textbf{Temporal representations and evidence access.}
TimeAudio combines temporal markers, absolute time-aware encoding, and segment-level token merging~\cite{wang2026timeaudio}. TimePro-RL interleaves timestamp embeddings with audio features and applies GRPO with an adaptive temporal reward~\cite{shi2026timepro}. Related video methods encode time through temporal separators in TimeMarker~\cite{chen2024timemarker} or a distribution-based time encoder--decoder in DisTime~\cite{zeng2025distime}. For long audio, LAT-Audio constructs a global timeline and accesses local segments through TWA-CoT~\cite{shao2026lataudio}; LA-RAG retrieves timestamped event records from a structured database~\cite{hegde2026laragqa}. These methods improve temporal representations, evidence access, or final-answer alignment. TimePro-RL optimizes the temporal accuracy of sampled answers; our additional target is the change from each draft to its revised answer. \method{} retains the backbone and performs count/boundary diagnosis and interval revision within one response, without a retrieval system or local-audio tool loop.

\noindent\textbf{Reasoning and reflection training.}
Audio Flamingo Sound-CoT~\cite{kong2025soundcot} constructs explicit reasoning chains through interactions between language and audio models, then fine-tunes for multiple-choice sound question answering and classification. SRPO~\cite{wan2025srpo} combines reflection-oriented SFT with GRPO and scores correctness transitions, reflection format, and response length. \method{} instead expresses temporal reflection through count/boundary actions and revised intervals.

\noindent\textbf{Evaluation scope.}
MMAU and MMAR evaluate audio understanding and multi-step reasoning~\cite{sakshi2024mmau,ma2025mmar}; event-presence sensitivity analysis tests detection and false alarms in multi-event scenes~\cite{lee2026multievent}. TAG-Bench instead measures timestamp localization and occurrence enumeration~\cite{dai2026tagbench}. We complement these final-output metrics with count- and width-matched random-draft comparisons and draft-to-final quality changes, separately testing whether drafts encode temporal information and whether revision improves localization.

\section{Method}
\label{sec:method}
\method{} learns temporal self-correction in two stages (Fig.~\ref{fig:pipeline}). We denote final-answer-only fine-tuning by \emph{SFT}, trajectory-based fine-tuning by \emph{\recot-SFT}, and its RL continuation by \emph{\recot-SFT-RL}, our final model. We first formalize the task, then describe the protocol, data construction, and reflection objective.

\begin{table*}[t]
\centering
\caption{Main grounding results: mIoU and recall in percent. Avg. is the unweighted mean of Clotho, LAT, and TAG-Bench All mIoU. Bold and underlined values denote the best and second-best results, respectively.}
\label{tab:main}
\small
\setlength{\tabcolsep}{8.5pt}
\begin{tabular}{lcccccccc}
\toprule
\multirow{2}{*}{Model} & \multirow{2}{*}{Clotho} & \multirow{2}{*}{LAT} & \multicolumn{3}{c}{TAG-Bench mIoU} & \multicolumn{2}{c}{TAG-Bench recall} & \multirow{2}{*}{Avg.}\\
\cmidrule(lr){4-6}\cmidrule(lr){7-8}
& & & All & Single & Multi & R@.5 & R@.7 & \\
\midrule
MOSS-Audio-8B-Thinking~\cite{openmoss2026audio} & 54.57 & 18.56 & 29.6 & 31.4 & 23.6 & 25.5 & 18.2 & 34.24\\
Gemini-3.1-Pro~\cite{gemini31} & 70.35 & \textbf{30.8} & 26.4 & 27.9 & 20.9 & 24.0 & 13.7 & 42.52\\
FireRedAudio~\cite{fireredaudio} & 54.70 & 27.37 & 31.2 & 34.1 & 21.0 & 30.8 & 21.5 & 37.76\\
\midrule
SFT & 84.01 & 28.46 & 36.55 & 38.64 & 29.20 & 31.66 & 23.20 & 49.67\\
\recot-SFT & \textbf{86.11} & 28.32 & \underline{38.26} & \underline{40.68} & \underline{29.73} & \underline{33.37} & \underline{23.71} & \underline{50.90}\\
\rowcolor{laglightblue}
\recot-SFT-RL ($\alpha=0.3$) & \underline{85.99} & \underline{28.55} & \textbf{38.50} & \textbf{40.87} & \textbf{30.13} & \textbf{34.23} & \textbf{23.89} & \textbf{51.01}\\
\bottomrule
\vspace{-0.7cm}
\end{tabular}
\end{table*}

\subsection{Task Formulation}
\label{ssec:task}
Given an audio recording $x$ lasting $D$ seconds and a natural-language query $q$, we predict all intervals matching $q$. Let $G=\{[s_j,e_j]\}_{j=1}^{M}$ and $P=\{[\hat{s}_k,\hat{e}_k]\}_{k=1}^{N}$ be the reference and predicted interval sets, containing $M$ and $N$ intervals, respectively. Here, $(s_j,e_j)$ and $(\hat{s}_k,\hat{e}_k)$ are the start/end times of reference interval $j$ and prediction $k$. Times are in seconds from audio onset, with $0\leq s<e\leq D$ for any valid interval $[s,e]$. Outputs use semicolon-separated \texttt{start\textasciitilde end} ranges.

Let $\mathcal U(\cdot)$ denote temporal union, merging overlapping intervals, and $|\cdot|$ denote duration in seconds. We measure intersection over union as
\begin{equation}
\operatorname{IoU}(P,G)=
\frac{|\mathcal U(P)\cap\mathcal U(G)|}
{|\mathcal U(P)\cup\mathcal U(G)|}.
\label{eq:iou}
\end{equation}
Task metrics score final intervals; drafts enter reward computation and process diagnostics.

\subsection{Reflection Chain-of-Thought}
\label{ssec:recot}
\recot{} uses a fixed draft--diagnosis--revision response:
\par\smallskip
{\centering
\begin{adjustbox}{max width=\columnwidth}
\begin{tabular}{@{}l@{}}
\texttt{<think><draft>10\textasciitilde14</draft><count>keep</count>}\\
\texttt{<boundary>right</boundary></think>12.26\textasciitilde16.47}
\end{tabular}
\end{adjustbox}
\par}\smallskip
The \texttt{draft} contains initial intervals. The \texttt{count} action is one of \{keep, add, remove\}, and the \texttt{boundary} action is one of \{keep, left, right, expand, shrink, adjust\}. Intervals after \texttt{</think>} form the revised answer.
Multi-target responses use semicolon-separated draft/final intervals with one count/boundary action pair for the entire set.

Count and boundary actions distinguish occurrence mismatch from temporal misalignment at the set level. Boundary \texttt{keep} denotes agreement under the labeling thresholds, not identical timestamps; \texttt{adjust} covers residual cases. Fixed fields make drafts and revisions deterministically parseable, enabling a shared quality function to evaluate both without a free-text judge.

\subsection{Training Data Construction}
\label{ssec:data}
\noindent\textbf{Ground-truth-conditioned LLM generation.}
Sources include our in-house time-aligned speech corpus, long-form recordings, AudioSet-Strong~\cite{hershey2021strong}, and Clotho-derived grounding examples~\cite{drossos2020clotho}. As shown in Fig.~\ref{fig:pipeline}, an LLM first constructs candidate \recot{} trajectories from queries and pre-annotated reference intervals: initial predictions, count/boundary diagnoses, and revised answers. The references anchor the final targets; the generated drafts and diagnoses provide the intermediate correction cases.

\noindent\textbf{Rule-based post-processing and balancing.}
We then reconcile generated intervals and action labels using deterministic rules, preserving reference intervals as final targets. These rules standardize supervision rather than generate trajectories from scratch. Count labels follow interval cardinality. Boundary \texttt{keep} requires equal counts, union IoU $\geq0.70$, and matched IoU $\geq0.60$. Otherwise, duration-weighted center offsets assign \texttt{left}/\texttt{right}, with a tolerance of one quarter of median target width, clipped to 0.2--5\,s. If neither shift applies, covered-duration differences above $\max(0.2\,\mathrm{s},0.15|\mathcal U(G)|)$ assign \texttt{expand}/\texttt{shrink}; remaining cases use \texttt{adjust}. These scale-dependent tolerances standardize offline labels, not test-time acceptance or RL action rewards. Syntax and consistency checks yield 275,000 trajectories balanced across correction cases and all action types. 

\noindent\textbf{The 5K RL subset.}
We select 5,000 SFT examples balanced by query type, duration, and target multiplicity. Audio, queries, durations, and references are retained; offline assistant trajectories are removed. Policy rollouts generate fresh drafts, actions, and final intervals; references provide rewards, not policy inputs.

\vspace{-0.1cm}
\subsection{Reflection-aware GRPO}
\label{ssec:grpo}
We sample eight responses per prompt and normalize rewards within each group using GRPO~\cite{shao2024deepseekmath}. Drawing on TAG-Bench's overlap, midpoint-error, and count metrics~\cite{dai2026tagbench}, we score drafts and final answers with
\begin{equation}
Q(P,G)=I_u + I_m + r_{\mathrm{mae}} + c_{\mathrm{soft}},
\label{eq:quality}
\end{equation}
Here $I_u$ is union IoU in Eq.~\ref{eq:iou}; $I_m$ is the maximum one-to-one sum of pairwise IoUs divided by $\max(N,M)$. Only union-based measures merge overlaps; matching and counting retain individual intervals, including duplicates. Midpoint MAE averages each reference midpoint's distance to its nearest predicted midpoint, allowing prediction reuse. It has no separate unmatched penalty; $I_m$ and count agreement address cardinality mismatch. We use $r_{\mathrm{mae}}=1/(1+\mathrm{MAE}_{\mathrm{mid}}/s)$, with $s=\max(1\,\mathrm{s},0.05D)$, and $c_{\mathrm{soft}}=\max(0,1-|N-M|/\max(N,M))$. All four components and $Q$ lie in $[0,1]$ and are zero if either set is empty. The response/reference grammar requires nonempty, start-ordered interval lists. The final reward is $R_{\mathrm{fin}}=Q(P_{\mathrm{fin}},G)$.

Let $Q_{\mathrm{dft}}=Q(P_{\mathrm{dft}},G)$ and $Q_{\mathrm{fin}}=Q(P_{\mathrm{fin}},G)$ for draft/final sets. With $\Delta Q=Q_{\mathrm{fin}}-Q_{\mathrm{dft}}$, we use exact comparisons:
\begin{equation}
R_{\mathrm{ref}}=
\begin{cases}
 0.50, & \Delta Q>0,\\
 0.25, & \Delta Q=0,\ Q_{\mathrm{dft}}>0.75,\\
-0.10, & \Delta Q=0,\ Q_{\mathrm{dft}}\leq0.75,\\
-0.25, & \Delta Q<0.
\end{cases}
\label{eq:rref}
\end{equation}
Equation~\ref{eq:rref} rewards improvement and preservation of high-quality drafts while penalizing missed or harmful revisions. We combine this signal with final quality:
\begin{equation}
R=\alpha R_{\mathrm{fin}}+(1-\alpha)R_{\mathrm{ref}}.
\label{eq:reward}
\end{equation}

We evaluate $\alpha\in\{0,0.3,0.5,0.7,1\}$, using 0.3 for the final model. Both rewards are set to $-1$ for malformed, empty, or truncated responses, or responses that fail temporal boundary validation. Diagnosis compares declared actions with labels derived from draft/reference intervals; execution agreement instead uses draft/final intervals. Both average count and boundary correctness and are diagnostics, not rewards.


\section{Experiments}
\label{sec:experiments}
\subsection{Setup}



To assess grounding performance against open-source and commercial baselines, we compare \method{} with MOSS-Audio-8B-Thinking~\cite{openmoss2026audio}, FireRedAudio~\cite{fireredaudio}, and Gemini-3.1-Pro~\cite{gemini31}. The first two are open-source audio models; the third is a commercial multimodal model.

SFT and \recot-SFT share Qwen3-Omni-30B-A3B-Instruct initialization and source examples; their supervision differs only in response structure, as defined in Section~\ref{sec:method}. Both train for 500 steps with global batch size 32, maximum sequence length 16,384, and initial learning rate $5\times10^{-6}$. Audio/vision towers and aligners are frozen. \recot-SFT-RL continues from the \recot-SFT checkpoint.

The five reward-weight configurations share the \recot-SFT checkpoint, balanced 5K data, and one-epoch schedule (approximately 300 steps), varying only $\alpha$ in Eq.~\ref{eq:reward}. We use attention-only LoRA (rank 8, scale 16), eight responses per prompt, 16 GPUs, gradient accumulation 8, learning rate $5\times10^{-7}$, KL coefficient 0.03, clipping 0.2, temperature 1.0, and top-$p$ 0.95.

Evaluation spans three settings: Clotho-derived grounding~\cite{drossos2020clotho} uses short environmental-audio clips (5,472 queries); LAT-Audio-TAG-Test~\cite{shao2026lataudio} tests long-form audio with semantic queries (1,006); TAG-Bench~\cite{dai2026tagbench} covers diverse query types and recurring targets (1,750; 1,363 single-target, 387 multi-target). Decoding uses temperature 0 and at most 384 new tokens. Final intervals determine mean union IoU and Recall@$\tau$ (fraction with IoU $\geq\tau$); invalid answers count as empty predictions. Table~\ref{tab:main}'s Avg. weights datasets equally; Overall weights all 8,228 queries equally. Process diagnostics use the same full $Q$. Results are single-run estimates.

\subsection{Main Results}
\label{ssec:main}
\recot-SFT improved Avg. mIoU from 49.67\% to 50.90\% over SFT (Table~\ref{tab:main}); its query-weighted gain was 1.74 points before rounding. TAG-Bench overall mIoU rose by 1.71 points, with gains on both target groups and recall thresholds.

RL further raised Avg. mIoU to 51.01\%. Its gains over \recot-SFT were 0.40 points in TAG-Bench multi-target mIoU and 0.86/0.18 points in R@0.5/R@0.7. LAT mIoU rose by 0.23 points. The 0.12-point decrease on Clotho, which contains 66.5\% of queries, partly offset these gains in query-weighted mIoU (68.86\% to 68.87\%). Thus, RL added multi-target localization and recall gains while retaining overall accuracy.

\method{} outperformed the evaluated open-source MOSS-Audio and FireRedAudio models and commercial Gemini-3.1-Pro on all five TAG-Bench metrics and Avg. mIoU. The Avg. margin over the strongest external baseline, Gemini-3.1-Pro, was 8.49 points. TAG-Bench overall and multi-target margins over the strongest external scores were 7.30 and 6.53 points. The matched SFT comparison isolates our training change.

\subsection{Draft Quality and Reflection Effectiveness}
\label{ssec:diag}
Table~\ref{tab:reflect} separates draft informativeness from revision gain. \recot-SFT drafts scored above a count- and width-matched random-position baseline (32 placements) on every dataset, by 0.2428 in aggregate $Q$. Revision raised $Q$ from 0.6487 to 0.7526 and mIoU from 55.41\% to 68.86\%, complementing the final-answer comparison with SFT.

LAT showed little change from draft to final: mIoU rose from 28.31\% to 28.32\%, while $Q$ changed from 0.4997 to 0.4963. Its draft already reached nearly the final localization level and scored 0.2276 above random placement. This pattern likely reflects that long-form semantic queries depend more on identifying the relevant evidence across the recording; once the draft has selected an interval, revising counts and boundaries without additional evidence yields limited gains.

RL improved TAG-Bench multi-target draft and final mIoU from 25.05\% to 25.56\% and from 29.73\% to 30.13\%, respectively. Aggregate revision gain remained close to \recot-SFT (0.1038 vs.\ 0.1035). Its benefit is better final localization, not a larger average correction step.

\begin{table}[t]
\centering
\caption{\recot-SFT diagnostics. $Q_d/Q_r/Q_f$: draft/count-width-matched random/final quality ($\times100$); $\Delta_r=Q_d-Q_r$, $\Delta_f=Q_f-Q_d$. mIoU$_d$/mIoU$_f$: draft/final mIoU (\%). Differences precede rounding.}
\label{tab:reflect}
\small
\setlength{\tabcolsep}{2.1pt}
\begin{tabular}{lccccccc}
\toprule
& \multicolumn{3}{c}{Draft} & \multicolumn{2}{c}{Reflection gain} & \multicolumn{2}{c}{mIoU (\%)}\\
\cmidrule(lr){2-4}\cmidrule(lr){5-6}\cmidrule(lr){7-8}
Dataset & $Q_d$ & $Q_r$ & $\Delta_r$ & $Q_f$ & $\Delta_f$ & mIoU$_d$ & mIoU$_f$\\
\midrule
Overall & 64.87 & 40.60 & +24.28 & 75.26 & +10.38 & 55.41 & 68.86\\
Clotho & 74.11 & 46.18 & +27.93 & 87.81 & +13.70 & 68.44 & 86.11\\
LAT-Audio & 49.97 & 27.20 & +22.76 & 49.63 & $-0.34$ & 28.31 & 28.32\\
TAG-Bench & 44.57 & 30.83 & +13.74 & 51.32 & +6.75 & 29.12 & 38.26\\
\bottomrule
\vspace{-0.5cm}
\end{tabular}
\end{table}

\subsection{Reward Design Ablations}
\label{ssec:ablation}
\begin{figure}[t]
\centering
\includegraphics[width=\columnwidth]{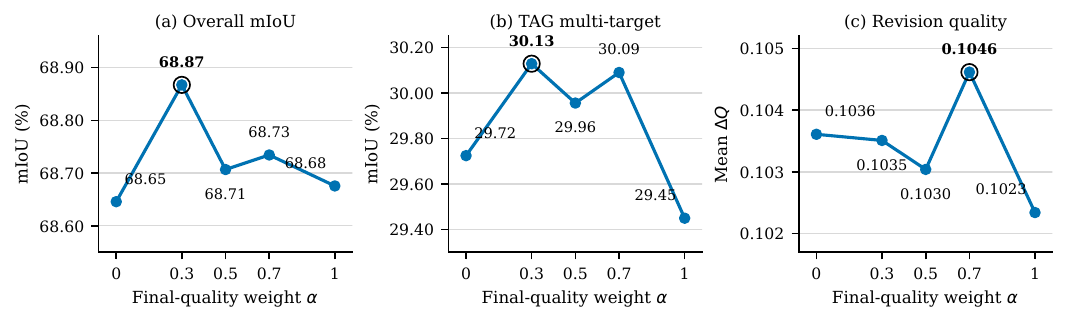}
\vspace{-0.5cm}
\caption{Reward-weight sensitivity under matched initialization, 5K data, and schedule (single run). Each panel sweeps $\alpha$, the weight on final quality, while $1-\alpha$ weights reflection. Panel (c) averages revision gain ($\Delta Q$) over 8,228 queries. Localization peaks at $\alpha=0.3$ and revision gain at $\alpha=0.7$, separating correction magnitude from final accuracy.}
\label{fig:alpha}
\end{figure}

\noindent\textbf{Reward mixing.}
Figure~\ref{fig:alpha} tests the value of revision feedback. Every mixed reward achieved higher multi-target mIoU than either endpoint. At $\alpha=0.3$, gains over reflection-only and final-only were 0.41 and 0.68 points, supporting the revision signal made measurable by \recot{}. This setting maximized Overall and multi-target mIoU; mean revision gain peaked at $\alpha=0.7$, separating correction magnitude from final accuracy.


\noindent\textbf{Quality components.}
\label{ssec:q_ablation}
Table~\ref{tab:q_ablation} tests the four criteria in Eq.~\ref{eq:quality}, renormalizing remaining weights after each removal. Variants share SFT initialization, 5K data, one-epoch update schedule, and $\alpha=0.3$; both rewards use the variant $Q$, with thresholds and parsing unchanged.

Full $Q$ achieved the highest Overall and TAG-Bench overall mIoU: removing any component reduced them by 0.23--0.44 and 0.53--1.04 points, respectively, and reduced both recall scores. Full $Q$ also had the lowest harmful-revision rate under the shared evaluation $Q$ (13.98\% vs.\ 14.22--14.69\%). Removing $c_{\mathrm{soft}}$ improved multi-target mIoU; removing $I_m$ improved LAT mIoU (29.05\% vs.\ 28.55\%). The full objective therefore benefits overall grounding, while individual subsets favor different criteria.

\begin{table}[!ht]
\centering
\caption{Q-component ablation (\%, $\uparrow$), using the symbols in Eq.~\ref{eq:quality}. Overall is query-weighted mIoU over 8,228 queries. TAG All/Multi are mIoU; recall uses all 1,750 TAG-Bench queries. Bold/underline mark best/second-best.}
\label{tab:q_ablation}
\small
\setlength{\tabcolsep}{2.8pt}
\begin{tabular}{lccccc}
\toprule
\multirow{2}{*}{Variant} & \multirow{2}{*}{Overall} & \multicolumn{4}{c}{TAG-Bench}\\
\cmidrule(lr){3-6}
& & All & Multi & R@.5 & R@.7\\
\midrule
Full $Q$ & \textbf{68.87} & \textbf{38.50} & \underline{30.13} & \textbf{34.23} & \textbf{23.89}\\
w/o $I_u$ & 68.50 & 37.45 & 29.58 & 32.34 & 22.57\\
w/o $I_m$ & 68.50 & 37.81 & 29.56 & 32.91 & 22.69\\
w/o $r_{\mathrm{mae}}$ & 68.43 & \underline{37.97} & 29.92 & 33.26 & \underline{23.09}\\
w/o $c_{\mathrm{soft}}$ & \underline{68.64} & 37.91 & \textbf{30.51} & \underline{33.37} & 22.74\\
\bottomrule
\vspace{-0.7cm}
\end{tabular}
\end{table}

\section{Conclusion}
\label{sec:conclusion}
\method{} makes temporal self-correction measurable through structured drafts, actions, and revisions. \recot-SFT learns informative drafts and effective aggregate correction; RL adds multi-target localization and recall gains. Reward ablations support combining final quality and revision feedback with complementary temporal criteria. At 38.50\% TAG-Bench mIoU and 51.01\% Avg. mIoU, our final model outperforms the evaluated open-source MOSS-Audio and FireRedAudio models and commercial Gemini-3.1-Pro, supporting structured temporal reflection as an effective post-training approach.

\clearpage
\footnotesize
\bibliographystyle{IEEEbib}
\bibliography{strings,refs}
\end{document}